\documentclass[10pt]{article}
\usepackage[english]{babel}
\usepackage{amsmath,amsthm,amssymb, enumerate}
\usepackage[utf8]{inputenc}
\usepackage[margin=3cm]{geometry}
\usepackage{charter}
\usepackage{tikz}
\usepackage[utf8]{inputenc} 
\usepackage[T1]{fontenc}    
\usepackage{url}            
\usepackage{booktabs}       
\usepackage{amsfonts}       
\usepackage{nicefrac}       
\usepackage{microtype}      
\usepackage{amsfonts}  
\usepackage{graphicx}
\usepackage{fullpage}

\usepackage{graphicx, float}
\usepackage{subcaption}
\usepackage{url}
\usepackage{epsfig}
\usepackage{color}
\usepackage{refcount}
\usepackage{verbatim}
\usepackage{enumerate}
\usepackage{multirow}
\usepackage{framed}
\usepackage{spverbatim}
\usepackage{ourStyle}
\usepackage{tikz}
\usepackage{color}
\definecolor{clemson-orange}{RGB}{234,106,32}
\definecolor{chicago-maroon}{RGB}{128,0,0}
\definecolor{cincinnati-red}{RGB}{190,0,0}
\definecolor{soft-cyan}{RGB}{68,85,90}
\usepackage{fullpage}
\usepackage{authblk}
\usepackage{mathtools}
\usepackage[parfill]{parskip}

\makeatletter
\@ifundefined{theorem}{%

}{}
\@ifundefined{conjecture}{%

}{}
\@ifundefined{lemma}{%

}{}
\@ifundefined{claim}{%

}{}
\@ifundefined{observation}{%

}{}
\@ifundefined{definition}{%

}{}
\@ifundefined{question}{%

}{}
\@ifundefined{corollary}{%

}{}
\makeatother

\title{Tipping Points for Norm Change in Human Cultures}
\author{Soham De\textsuperscript{1,2}, Dana S. Nau\textsuperscript{1,2}, Xinyue Pan\textsuperscript{3} and Michele J. Gelfand\textsuperscript{3}}
\affil{\textsuperscript{1}Department of Computer Science, \textsuperscript{2}Institute for Systems Research, \\
\textsuperscript{3}Department of Psychology,
\textsuperscript{1,2,3}University of Maryland, College Park \\ \vspace{2mm} \texttt{\{sohamde, nau\}@cs.umd.edu, \{xypan, mgelfand\}@umd.edu} \\ \vspace{2mm}
{\small 
The final authenticated publication is available online at \texttt{https://doi.org/10.1007/978-3-319-93372-6\_7}}}
 \date{}

\begin{document}

\maketitle

\begin{abstract}
Humans interact with each other on a daily basis by developing and maintaining various social norms and it is critical to form a deeper understanding of how such norms develop, how they change, and how fast they change. In this work, we develop an evolutionary game-theoretic model based on research in cultural psychology that shows that humans in various cultures differ in their tendencies to conform with those around them. Using this model, we analyze the evolutionary relationships between the tendency to conform and how quickly a population reacts when conditions make a change in norm desirable. Our analysis identifies conditions when a tipping point is reached in a population, causing norms to change rapidly.
\end{abstract}

\section{Introduction}
\label{sec:intro}
Social norms are critical in enabling human populations across the world to coordinate and accomplish different tasks. The \emph{strength} of these social norms, however, differs widely around the globe, as has been established by past neuroscience, field and experimental research 
\cite{gelfand2011differences, harrington2014tightness, harrington2015culture, mu2015culture, roos2015societal, de2017understanding, gelfand2017strength}. 
Some cultures are said to be \emph{tight}, with strong social norms, typically characterized by high degrees of norm adherence and strict punishment directed towards norm-violators. Other cultures are said to be \emph{loose}, with weaker norms characterized by a higher acceptance of deviant behavior \cite{gelfand2011differences, harrington2014tightness, roos2015societal}.

Tightness-looseness is a dynamic construct, yet to date, there has been little research on the evolutionary processes that lead to \emph{changes} in societal norms, the \emph{rate} at which such changes occurs, and how these processes \emph{vary} across different cultures. In this paper, we aim to study how cultural differences in the way humans interact and influence each other heavily influence how societal norms are established and the rate at which they change across the world. We use evolutionary game theory (EGT) to examine the causal relationship between an individual's tendency to conform with those around them and the rate at which norms are changed in different cultures (see Section \ref{sec:related} for a brief discussion on EGT). More specifically, our primary contributions in this paper are as follows:
\begin{itemize}
\item Drawing on recent research in cultural psychology, we propose a game-theoretic model of a culture based on the tendency of an individual to conform with others, vs.~being more individualistic in their behavior.
\item Using this model, we provide conditions under which a population is open to changing the current norm in a society, depending on the pressure of conformity and the abruptness of the boundary between following and violating the norm.
\item Finally, we analyze the \emph{rate} at which such norm changes occur.
We show that tighter cultures are more likely to be initially resistant to norm changes compared to looser cultures. Further, we analyze conditions under which tighter cultures sometimes reach a \emph{tipping point}, where a large proportion of the population suddenly switch to a new norm.
\end{itemize}




The rest of the paper is organized as follows. Section \ref{sec:related} provides background and related work. We introduce our proposed model in Section \ref{sec:proposed} and study the rate of norm change in Section \ref{sec:rate}. In Section \ref{sec:discussion} we discuss the significance of our results.

\section{Background and Related Work}
\label{sec:related}
Evolutionary Game Theory (EGT) was initially proposed to model biological evolution \cite{hofbauer2003evolutionary, smith1982evolution, weibull1997evolutionary}, but has been increasingly used to study human cultural evolution. The idea is to represent an interaction among individuals as a normal-form game, where individuals can use different strategies. The game's payoffs represent an individual's evolutionary fitness. In EGT models of cultural evolution, biological reproduction represents \emph{social learning}: individuals are more likely to adopt strategies from others that produce high fitness and thus strategies that lead to higher fitness become more prevalent over time.
While such models use highly simplified abstractions of complex human interactions, they aim to capture the essential nature of the interactions of interest, and thus have been increasingly used to study a wide variety of social and cultural phenomena,
(see \cite{de2016using} for a brief tutorial on how such models are generally set up)
such as, cooperation and altruism \cite{bowles2004evolution, hamilton1981evolution, nowak2006five, nowak1992tit, riolo2001evolution}, punishment \cite{boyd2003evolution, brandt2003punishment, brandt2006punishing, rand2011evolution, roos2014high}, trust and reputation \cite{Fang2002, hauert2010replicator, 10.1007/11429760_1}, ethnocentrism \cite{de2015inevitability, hammond2006evolution, hartshorn2013evolutionary}, etc.

There has been widespread interest in studying the \emph{emergence} of social norms in a population both from an evolutionary perspective \cite{young2001individual, hechter2001social, merton1938science, bicchieri2005grammar, helbing2014conditions}, as well as in empirical research \cite{judd2010behavioral, kearns2009behavioral, centola2015spontaneous}. There has been, however, much less work done on understanding the processes that lead to \emph{change} in an already established norm in a population.
A related concept, the propagation of information in social networks, has been well-studied (see \cite{chen2013information, jackson2010social, easley2010networks}  
for an overview), but these works typically do not account for the differences in how individuals interact and influence each other in different cultures. 
Data science approaches have also explored this question, however, it is very challenging to separate out the various confounding factors (such as institutional influence) to establish clear causal relationships \cite{zhang2015creates, lehmann2012dynamical, kooti2012emergence, lin2013bigbirds}. Finally, note that \cite{henrich2001cultural} and \cite{de2017understanding} have previously studied the processes of norm change in societies, and our work extends these studies in important directions. Using a more general model of conformist transmission that depends on the degree to which there is an abrupt boundary between following and violating the norm, we study the \emph{speed} of norm change in different cultures, and show how \emph{tipping points} during such norm change depends critically on the model of conformist transmission.

\section{Proposed Evolutionary Game-Theoretic Model}
\label{sec:proposed}
Research in cross-cultural psychology has established that the strength of social norms varies considerably across cultures. Further, using historical and ecological data, it has been shown that this is related to the degree of threat that populations face \cite{gelfand2011differences, harrington2014tightness}. Stronger norms and sanctions are needed in high-threat situations to coordinate and survive, leading to tighter cultures. By contrast, populations that lack exposure to serious ecological threats can afford to have weaker norms and tolerance for deviance given that they have less need for coordinated social action (looser cultures). We now describe our model.

Consider an infinite, well-mixed population (i.e., each individual can interact with any other individual in the population) that evolves according to the well-known replicator dynamic \cite{hofbauer2003evolutionary}. For simplicity of presentation, suppose each agent may choose one of two possible actions: $A$ and $B$ (see Section \ref{sec:discussion} for a discussion on the assumptions used in our model). The two actions $A$ and $B$ correspond to possible norms that the society could settle on. Let $x_A$ and $x_B$ denote the proportions  of the population using actions $A$ and $B$ respectively, with $0 \le x_A, x_B \le 1$ and $x_A + x_B = 1$, and let $x = (x_A, x_B)$.
According to the replicator dynamic, the rate of change in the proportions of agents using each action is given by the following differential equation:
\begin{align}
 \dot{x}_i = x_i [ f_i(x) - \phi(x)], \label{replicator}
\end{align}
where $i \in \{A, B\}$, $\dot{x}_i = dx_i/dt$ (i.e., rate of change of $x_i$), $f_i(x)$ is the fitness of action $i$, and $\phi(x)$ denotes the average fitness of the population, i.e.:
$
\phi(x) = x_A f_A(x) + x_B f_B(x).
$
The replicator dynamic is based on the idea that the proportion of agents with a particular strategy increases when it achieves expected fitness higher than the average fitness, and vice versa.

\begin{figure}[t]
\centering
\includegraphics[width=0.28\textwidth]{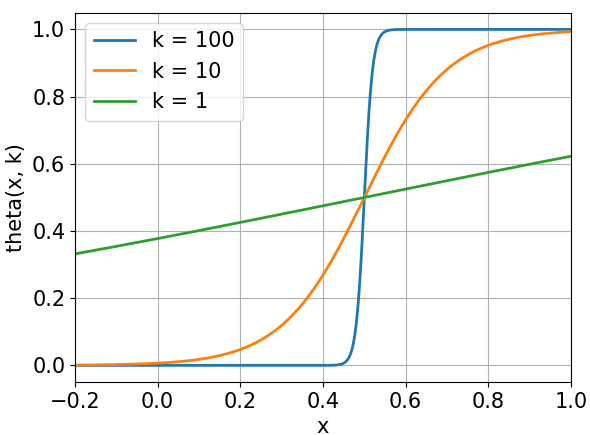}
\includegraphics[width=0.35\textwidth]{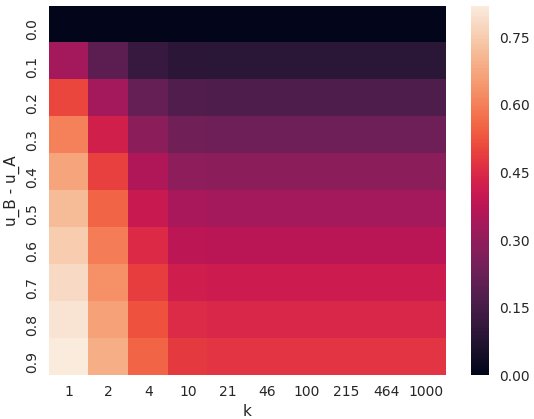}
\includegraphics[width=0.35\textwidth]{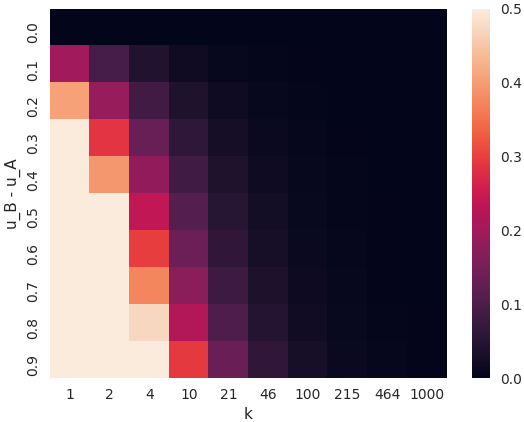}
\caption{\emph{Left:} Plot of \eqref{theta} for different values of $k$. \emph{Middle:} Heatmap of the right-hand side in \eqref{m_change}
 when $x_B=0.1$, for various $u_B - u_A$ and $k$ values. \emph{Right:} Heatmap of the right-hand side in \eqref{epsilon_rate}, for various $u_B - u_A$ and $k$ values. Best viewed in color.}
\label{theta_m_change}
\vspace{-4mm}
\end{figure}

Let $u_A$ and $u_B$ denote the payoffs associated with actions $A$ and $B$, where $0 < u_A, u_B < 1$ and $u_A + u_B = 1$. To define the fitness function $f_i$, we use the key insight that in loose cultures, individuals tend to choose the action that is most beneficial to them; but in tight cultures, individuals tend to conform to the same action that others use, even if a different action might be more beneficial to each individual. To model this mathematically, we let $f_i$ be a weighted combination of the payoff $u_i$ and an additional \emph{conformism fitness} measure $\theta_i$ that depends on whether the individual is conforming to others in the population. Let $m$ denote the parameter controlling the weighting between these two fitness measures, i.e., the amount of conformist transmission in a population. Thus, we define $f_i$ as:
\begin{align}
f_i(x, m) = (1-m) u_i + m \theta_i(x, k), \label{f}
\end{align}
where $0 \leq m \leq 1$, and we define the conformism fitness measure $\theta_i$ as:
\begin{align}
\theta_i(x, k) = \Big[ 1 + \exp \big(-k (x - 0.5)\big) \Big]^{-1}, \label{theta}
\end{align}
where $k > 0$. Note that we can vary the behavior of the conformism fitness measure $\theta_i$ using the parameter $k$ (see Figure \ref{theta_m_change}). For example, when $k$ is large, $\theta_i$ is close to a step function where there is an abrupt boundary between following and violating the norm and agents have a non-zero conformism fitness only if they conform with the majority action:
\begin{align*}
\theta^\infty_i(x) = \lim_{k \rightarrow \infty} \theta_i(x, k) &= \begin{cases}
	0,& \text{if } x_i < 0.5;\\
	0.5,& \text{if } x_i = 0.5;\\
	1,& \text{if } x_i > 0.5.\\
\end{cases}
\end{align*}
Note that with no conformism whatsoever ($m=0$), each action's fitness depends solely on its payoff, i.e., typical of a very loose culture.
On the other hand, with 100\% conformist transmission ($m=1$), $i$'s fitness depends solely on the conformism fitness measure (for the case of $\theta^\infty_i$, this means that $i$'s fitness depends solely on whether $i$ is in the majority or the minority of the population). This is more indicative of a very tight culture. For simplicity, for the rest of the paper, we denote $\theta_i(x, k)$ as $\theta_i$ and $\theta^\infty_i(x)$ as $\theta^\infty_i$. 

\subsection{When does norm change occur?}
\label{sec:change}
Suppose norm $B$ has a higher utility compared to $A$, i.e., $u_B > u_A$. We are interested in analyzing the conditions for which a population shifts from norm $A$ to $B$ (norm change). We can re-write the average fitness to be:
\begin{align}
\phi(x) = (1-m) (x_A u_A + x_B u_B) + m(x_A \theta_A + x_B \theta_B). \label{phi}
\end{align}
We are interested in anlayzing the rate of change in the proportion of $B$ individuals. From (\ref{replicator}), (\ref{f}), and (\ref{phi}), we get:
\begin{align}
\dot{x}_B &=
 x_B (1-x_B) \big[ (1-m) (u_B - u_A) + m ( \theta_B - \theta_A)\big]. \label{change}
\end{align}
Note that $\theta_B \ge \theta_A$ when $x_B \ge 0.5$. Since $u_B > u_A$, we see that $\dot{x}_B > 0$, i.e., $x_B$ will converge to 1 ($\lim_{t\rightarrow \infty} x_B = 1$) when $x_B \ge 0.5$. If $x_A > x_B$, i.e., if the current norm in the population is $A$, norm change takes place only if:
\begin{align}
m < \frac{u_B - u_A}{(u_B - u_A) + (\theta_A - \theta_B)}. \label{m_change}
\end{align}
Thus, norm change takes place only if the population is loose enough, while tighter cultures are more resistant to change.
Further, note that $\theta^\infty_A - \theta^\infty_B = 1$ when $x_A > x_B$. Thus, when the conformist fitness measure is a step-function $\theta^\infty_i$, \eqref{m_change} becomes: $m < (u_B - u_A)/(u_B - u_A + 1) < 0.5$. Thus, for $\theta^\infty_i$, norm change occurs only if individuals in a population weigh their individual payoff more than whether they conform with others. Figure \ref{theta_m_change} (middle) shows a heatmap of how condition \eqref{m_change} varies with $u_B - u_A$ and $k$ when $x_B = 0.1$. We see that the bound on $m$ increases as $u_B - u_A$ increases, i.e., a population becomes more likely to switch the norm. On increasing $k$, we see that the bound on $m$ decreases. This makes intuitive sense, since a higher $k$ makes the difference in conformist fitness between $A$ and $B$ clearer. Thus, in tight cultures, where people tend to agree more on what behaviors are appropriate vs.~inappropriate in different situations \cite{gelfand2011differences}, a higher $k$ would lead to more resistance to norm change.

\subsection{Rate of norm change in tight vs.~loose cultures}
\label{sec:rate}
We are now interested in studying the speed with which norms change in different populations. Consider two possible values of $m$, namely $m_1$ and $m_2$, with $m_2 > m_1$ (i.e., $m_2$ is a more conformist culture than $m_1$). Let the corresponding values of $\dot{x}_B$ be denoted by $\dot{x}_{B}^1$ and $\dot{x}_{B}^2$, respectively. Assume further that both $m_1$ and $m_2$ satisfy \eqref{m_change}, i.e., norm change takes place in both cultures. Analyzing the difference in the rates of change, from \eqref{change} we get:
\begin{align}
\dot{x}_{B}^2 - \dot{x}_{B}^1 =  x_B (1-x_B) (m_2 - m_1) \big[ ( \theta_B - \theta_A) - (u_B - u_A) \big].
\label{diff_rate}
\end{align}
Note that when $x_B \le 0.5$, $\theta_B - \theta_A \le 0$, which would mean: $\dot{x}_{B}^2 - \dot{x}_{B}^1 \le 0$, i.e., the more conformist culture would be slow to change initially. To analyze the case when $x_B > 0.5$, let's assume $x_B = 0.5 + \epsilon$, for $\epsilon > 0$. Thus, $x_A = 0.5 - \epsilon$. From \eqref{diff_rate}, we see that for $\dot{x}_{B}^2 - \dot{x}_{B}^1 > 0$, the following condition needs to hold:
\begin{align}
\epsilon > \big[ \ln(1 + u_B - u_A) - \ln(1 - (u_B - u_A)) \big]/k.
\label{epsilon_rate}
\end{align}
Figure \ref{theta_m_change} (right) plots a heatmap of how this bound varies with $u_B - u_A$ and $k$. We see that as $k$ increases, the point at which the more conformist culture starts changing faster moves closer to the point $x_B = 0.5$. Note that when $k \rightarrow \infty$, \eqref{epsilon_rate} reduces to $\epsilon > 0$, i.e., as soon as $x_B$ becomes a majority, greater conformism would produce a larger rate of change.
\begin{figure}[t]
\centering
\includegraphics[width=0.36\textwidth]{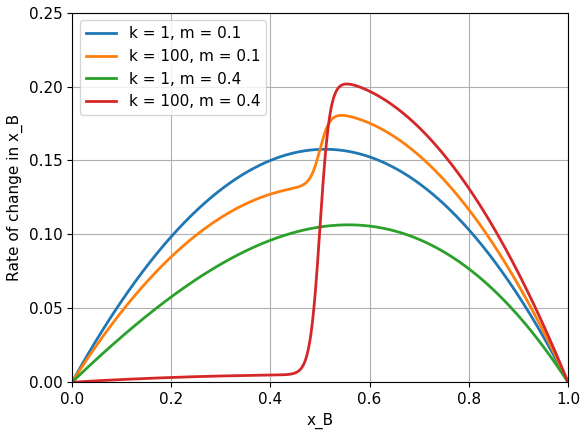}
\includegraphics[width=0.36\textwidth]{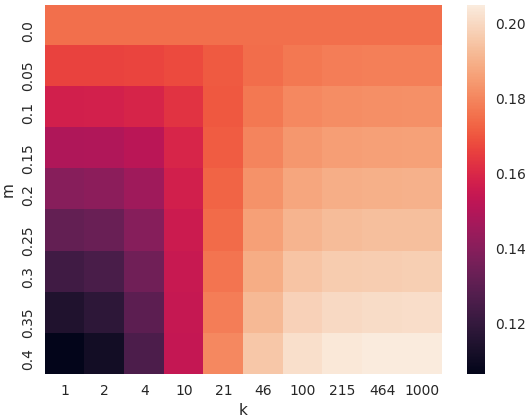}
\caption{\emph{Left:} Plot of \eqref{change} at $u_B - u_A = 0.7$. \emph{Right:} Heatmap of $\max_{x_B} \dot{x}_B$ for various $k$ and $m$ values, with $u_B - u_A = 0.7$. Best viewed in color.}
\label{rate_heatmaps}
\vspace{-4mm}
\end{figure}

We are also interested in studying how the rate of change $\dot{x}_B$ varies with $x_B$ as a population switches to norm $B$, and how this relates to different levels of conformism. We first look at the \emph{maximum} rate of change $\dot{x}_B^\text{max} = \max_{x_B} \dot{x}_B$ (we numerically calculate this given values for $k$, $m$ and $u_B - u_A$).
 Figure \ref{rate_heatmaps} (right) plots $\dot{x}_B^\text{max}$ for different $m$ and $k$ values, where we set $u_B - u_A = 0.7$. These values were chosen such that the norm changes from $A$ to $B$ for all the considered combinations (using the bounds from Section \ref{sec:change}). We see that when $k$ is low, lower conformism leads to higher $\dot{x}_B^\text{max}$. However, as $k$ increases (i.e., as $\theta_B$ approaches $\theta_B^\infty$), there is a clear transition, where more conformist cultures end up having a higher maximum rate of change.
 
This effect is clearer in the left plot of Figure \ref{rate_heatmaps}, where we show how $\dot{x}_B$ varies with $x_B$. We see that when $k$ is low, i.e., when there is no clear difference between $A$ and $B$ in its conformist fitness measure $\theta$, $\dot{x}_B$ changes slowly for both tight and loose cultures, with the loose culture having a higher rate of change.
With high $k$, however, we see that the tighter culture faces a \emph{tipping point}, resulting in a sudden increase in $\dot{x}_B$ with the tighter culture adopting a higher rate of change $\dot{x}_B$ than the loose culture.
Thus, using a more general model of conformist transmission over \cite{henrich2001cultural}, we find that the parameter $k$, which makes the difference in conformist fitness between following and violating the norm clearer, has a big influence on the pressure of conformity, and thus on the rate at which the norm changes in a society.
In summary, in a more conformist culture, initially peer pressure impedes the switch to the more beneficial norm $B$. But once enough of the population has switched, a tipping point is reached where peer pressure causes the rest of them to switch very rapidly.
\section{Discussion}
\label{sec:discussion}
This paper presents an EGT model that aims to investigate how the cultural dynamics of norm maintenance and norm-change differ across various cultures.
We show that tight cultures sometimes experience a tipping point for norm change while loose cultures typically face a more gradual change.
The results presented assume an infinite well-mixed population with two possible actions. We believe it would be relatively straightforward to extend our results for multiple actions. Assuming an infinite well-mixed population made our model mathematically tractable and to provide exact conditions for norm change. As future work, it would be interesting to extend this model to the finite population case, where interactions between individuals are dictated by a social network.

{\small
\bibliographystyle{alpha}
\bibliography{refs}
}
\end{document}